\documentclass[a4paper]{jpconf}
\usepackage{color,xcolor}
\usepackage{graphicx,epsfig,epsf,epstopdf}
\usepackage{multicol,setspace}
\usepackage{float}
\usepackage{caption}
\usepackage{amsmath,amssymb,amsfonts,fontenc}
\usepackage[titletoc,toc,page]{appendix}
\usepackage{lipsum}
\usepackage{lineno}
\def\beq{\begin{equation}}
\def\eeq{\end{equation}}
\def\beqn{\begin{eqnarray}}
\def\eeqn{\end{eqnarray}}
\begin{document}
\title{Many-Body Quantum Dynamics of a Bosonic Josephson Junction with a Finite-Range Interaction}

\author{Sudip Kumar Haldar and Ofir E Alon}

\address{Department of Mathematics and Haifa Research Center for Theoretical Physics and Astrophysics, University of Haifa, Haifa, Israel}

\ead{shaldar@campus.haifa.ac.il}

\begin{abstract}
	The out-of-equilibrium quantum dynamics of a Bose gas trapped in an asymmetric double well and interacting with a finite-range
	interaction has been studied in real space by solving the time-dependent many-body Schr\"odinger equation numerically
	accurately using the multiconfigurational time-dependent Hartree method for bosons (MCTDHB). 
	We have focused on the weakly interacting limit where the system is essentially condensed. 
	We have examined the impact of the range of the interaction on the
	dynamics of the system, both at the mean-field and many-body levels. Explicitly, we have studied 
	the maximal and the minimal values of the many-body position variance in each cycle of oscillation,
	and the overall pace of its growth. We find that the range of the interaction affects the dynamics of the system
	differently for the right well and the left well. We have also examined the infinite-particle limit and find that even there, 
	the impact of the range of the interaction can only be described by a many-body theory such as MCTDHB.
\end{abstract}

\section{Introduction}
\hspace*{0.5cm}
Following the experimental observation by the MIT group~\cite{Anderson}, the Bose-Einstein condensate (BEC) in an ultra-cold 
atomic gas has emerged as a popular test-bed for the interacting many-particle system~\cite{Dalfovo}. One of the main reasons for such 
huge interest in this system is its high degree of tunability and thereby its ability to simulate various condensed 
matter systems. One such well-studied system in condensed matter physics is the Josephson junction~\cite{Josephson}. An interacting Bose gas in a symmetric 
double well provides a paradigmatic model for the Josephson junction and is popularly known as the bosonic Josephson 
junction (BJJ)~\cite{Gati2007}. The BJJ has been extensively studied both theoretically and experimentally~[4-9].
Theoretically, two commonly used methods for the study of the BJJ dynamics are the mean-field method~\cite{Dalfovo} and the Bose-Hubbard model~\cite{Veksler}. 
While there are several features of the BJJ dynamics, such as the collapse and revival of the density oscillations~\cite{Milburn} and the 
fragmentation dynamics~\cite{Sakmann2014}, which can not be described by the mean-field method but can be described by the Bose-Hubbard model, 
the latter also fail to grasp the full many-body features~\cite{Sakmann2009}.

Recently, a numerically-exact many-body method has been used to study the BJJ dynamics for both the repulsive as well 
as attractive interactions~\cite{Ofir2008}. Specifically, fragmentation and the uncertainty product of the many-body position 
and momentum operators have been studied~\cite{MCHB, Klaiman2016}. It has also been shown that though there is a symmetry between the time 
evolution of the Bose-Hubbard dimer for the repulsive and attractive interaction of the same interaction strength, no such
symmetry exists for the many-body Hamiltonian~\cite{Sakmann2010}. Moreover, a universality of the fragmentation dynamics in the sense
that the condensates with different number of bosons but same interaction parameter fragment to the same degree has been found~\cite{Sakmann2014}.

Though all these works considered a contact $\delta-$interaction, a typical finite-range interaction has also been considered 
in a recent study and the impact of the range of the interaction on the BJJ dynamics has been examined~\cite{Sudip2018}. Such studies are
important in view of the recent experiments with the ultra-cold dipolar atoms where it has been clearly shown that both the 
short-range as well as the long-range terms are necessary to accurately describe the observed physics. BJJ dynamics was observed
to intricately depend on the competitive effects of the range of the interaction and the trapping potential. The tail of the interaction
potential was found to effectively enhance the effect of the repulsive interaction until the range became comparable with the length scale 
set by the trap.

Further, in a recent work, an asymmetric double well has also been considered and the effect of the loss of symmetry on the BJJ
dynamics has been thoroughly studied~\cite{Sudipasym}. It has been found that there is an overall suppression of the density oscillations 
in the asymmetric double well. Also, for a sufficiently strong interaction, the condensate becomes fragmented with time and the 
degree of fragmentation depends on the asymmetry as well as the initial well in a non-trivial manner. The universality of the fragmentation
is also observed in the asymmetric double well though the degree of fragmentation depends on the initial well.

In this work, we look back at the problem of the BJJ dynamics and combine the issues of the finite-range interaction and the asymmetric trap
in order to understand their combined impact on the BJJ dynamics. Hence, in this work, we consider the BJJ dynamics in an asymmetric double well
for a  typical finite-range interaction. We examine how the BJJ dynamics 
is affected by the tail of the interaction in presence of an asymmetry. We focus on the weakly interacting regime where the system is 
essentially fully condensed. Further, we examine the large particle limit and look for the combined effect of the range of interaction 
and the asymmetry of the trap at the
many-body level. We observe that even at the effectively infinite-particle limit, where the system is $100\%$ condensed, 
there are interesting effects at the
many-body level. The structure of the paper is as follows. In section~\ref{System}, 
we describe the system considered here. We present our results in Section~\ref{Results}
and summarize in Section~\ref{Summery}. The methodology and its accuracy are demonstrated in the appendices. 

\section{System}
\label{System}
\hspace*{0.5cm}
Here we are interested in a gas of $N$ interacting structureless bosons confined in a one-dimensional (1D) asymmetric double well $V_T(x)$. 
$V_T(x)$ is constructed by fusing two shifted harmonic wells $V_{L,R}(x)=\frac{1}{2}(x\pm2)^2+Cx$, i.e., 
\begin{equation}
V_T(x) = \left\{
\begin{matrix}
\frac{1}{2}(x + 2)^2 + Cx, \hspace*{1cm} x < -\frac{1}{2} \cr \cr
\frac{3}{2}(1-x^2) + Cx, \hspace*{1cm} |x| \le \frac{1}{2} \cr \cr
\frac{1}{2}(x - 2)^2 + Cx, \hspace*{1cm} x > \frac{1}{2} \cr  
\end{matrix}
\right.\,.
\end{equation}
The time period of Rabi oscillation $t_{Rabi}=\frac{2 \pi}{\Delta E}$ [$\Delta E$ being the energy gap between two lowest levels 
of $V_T(x)$] in $V_T(x)$ provides the natural time scale of the dynamics. In this work we consider only a small asymmetry $C=0.01$. For such a trap
$t_{Rabi}=102.0618$. The spatial separation $l=4$ between the two local minima of
the asymmetric double well trap sets the length scale for the system. The two-body inter-atomic interaction of strength $\lambda_0$ is modeled as 
$W(x_j-x_k)=  \frac{\lambda_0}{\sqrt{(|x_j-x_k|/D)^{6}+1}}$ of half-width $D$. $\lambda_0$ corresponds to 
the interaction parameter $\Lambda=\lambda_0(N-1)$. For a fixed $\Lambda$, the range of the interaction can be tuned by 
varying $D$ as a factor of the length scale $l$. Note that for $\frac{|x_j-x_k|}{D} \gg 1$, $W(x_j-x_k)$ behaves as the long-range dipolar interaction
while in the limit $\frac{|x_j-x_k|}{D} \rightarrow 0$, $W(x_j-x_k)$ reduces to a soft-core interaction having many similar effects of a short-range
$\delta$-interaction.

We characterize the dynamics of the system mainly by the time evolution of the 
many-body position variance of the system. The main reason for focusing on the time evolution of the many-body position variance is the 
recent observation that it can differ from the predictions of the mean-field Gross-Pitaevskii (GP) theory even when the system is fully condensed~\cite{Klaiman2015}.
Therefore, the many-body position variance can grasp those many-body effects which are otherwise washed out in the infinite-particle limit when the 
system is $100\%$ condensed and its density per particle and the energy per particle can be accurately described by the mean-field GP theory.
Further, we also study the growth in the number of depleted particles with time {to check the degree of condensation of the system.}

Given the many-body wavefunction $\Psi(t)$,
one can define the variance of the many-body position operator $\hat X=\sum_{j=1}^N \hat x_j$ as
\beqn\label{dis}
\frac{1}{N}\Delta_{\hat X}^2 &=& \frac{1}{N} 
\left[\langle\Psi|\hat X^2|\Psi\rangle - \langle\Psi|\hat X|\Psi\rangle^2\right] \nonumber \\
&=& \int dx \frac{\rho(x)}{N}x^2  - N \left[\int dx \frac{\rho(x)}{N}x\right]^2  \nonumber \\
&+& \sum_{jpkq} \frac{\rho_{jpkq}}{N(N-1)} \cdot (N-1) \int dx_1\, dx_2 \, 
\phi^{\ast{NO}}_j(x_1) \phi^{\ast{NO}}_p(x_2) \, x_1x_2 \, \phi^{NO}_k(x_1) \phi^{NO}_q(x_2). \
\eeqn
Here, $\rho(x)$ is the density of the system and is given by the diagonal element of the reduced one-body density matrix, i.e,
$\rho(x;t) \equiv \rho^{(1)}(x|x^{\prime}=x;t)$.  Similarly, $\rho_{ksql}$ is the matrix elements of the two-body reduced density matrix, viz., 
$\rho_{ksql}=\left<\Psi\left|b_k^\dag b_s^\dag b_q b_l\right|\Psi\right>$ where $b_k$ and $b_k^\dag$
are the bosonic annihilation and creation operators, respectively. The reduced one-body and two-body
density matrices $\rho^{(1)}(x_1|x_1^{\prime};t)$  and $\rho^{(2)}(x_{1},x_{2} \vert x_{1}^{\prime}, x_{2}^{\prime};t)$, respectively, are defined 
as
\beqn\label{1RDM}
\rho^{(1)}(x_1|x_1^{\prime};t) & = &
N \int d x_2 \ldots d x_N \, \Psi^\ast(x_1^{\prime},x_2,\ldots,x_N;t) \Psi(x_1,x_2,\ldots,x_N;t) \nonumber \\
& = &\sum_{j=1}^{M} n_j(t) \, \phi^{\ast{NO}}_j(x_1^{\prime},t)\phi^{NO}_j(x_1,t).
\eeqn

and 

\begin{equation}\label{2RDM}
\rho^{(2)}(x_{1},x_{2} \vert x_{1}^{\prime}, x_{2}^{\prime};t)= 
N(N-1)\int_{}^{}d x_{3} \ldots d x_{N} \Psi^{*}(x_{1}^{\prime},x_{2}^{\prime},x_{3},\ldots,x_{N};t) \Psi(x_{1},x_{2},x_{3}, \ldots, x_{N};t).
\end{equation} 

We point out that 
the eigenvalues and eigenvectors of $\rho^{(1)}(x_1|x_1^{\prime};t)$ are
the time-dependent natural occupation numbers $n_j(t)$ and the time-dependent natural orbitals $\phi^{NO}_j(x_1,t)$, respectively.
Further, $n_j(t)$ characterize the {(time varying)} degree of condensation 
in a system of interacting bosons \cite{PeO56} and satisfy $\sum_{j=1}^{M} n_j = N$ ($M$ is the number of single particle orbitals 
used to construct the many-boson wavefunction, see \ref{MCTDHB}).
If only one {macroscopic} eigenvalue 
$n_1(t) \approx {\mathcal O}(N)$ exists, the system is condensed \cite{PeO56} whereas if 
there are more than one {macroscopic} eigenvalues, the BEC
is said to be fragmented~[15,22-26].
For a condensed system, the microscopic occupations in all the higher orbitals $f = \sum_{j=2}^{M} \frac{n_j}{N}$
gives the depletion. On the other hand, the macroscopic occupation of a higher natural orbital $f=\frac{n_{j>1}}{N}$ gives the fragmentation.
Finally, we remark that substituting the many-body wavefunction $\Psi(t)$ in Eq.~(\ref{dis}) by the corresponding mean-field wavefunction, 
one can in study the variance of the many-body position operator at the mean-field level.

\section{Results}
\label{Results}
\hspace*{0.5cm}
In this section we describe the findings of our study of the 1D BJJ dynamics in an asymmetric trap. 
In an asymmetric double well trap, two wells are no longer equivalent - for the particular double well trap considered here, the left well $V_L(x)$
is lower than the right well $V_R(x)$. Accordingly, one can prepare the initial condensate state either in $V_L(x)$ or in $V_R(x)$. In this work, we start
the out-of-equilibrium dynamics of the system once from $V_L(x)$ and then from $V_R(x)$. As mentioned above, we consider 
the time evolution of the many-body position variance up to about $30$ Rabi periods, i.e., more than 3000 in absolute time units.
We consider systems of $N=100$ and $10000$ bosons interacting via a finite-range interaction of strength $\lambda_0$. 

As already mentioned, we are interested only in the very weakly interacting limit. Accordingly, we consider $\Lambda=0.01$. At such weakly interacting limit, 
the system is expected to be essentially condensed and, therefore, its out-of-equilibrium dynamics should be adequately described by the 
mean-field GP theory. So, first we consider the time evolution of the position variance at the mean-field level. 
The corresponding symmetric double well case, as discussed in Ref.~\cite{Sudip2018}, will serve as a reference for our study.

\begin{figure}[!ht]
\begin{center}
\includegraphics[width=0.32\linewidth, angle=270]{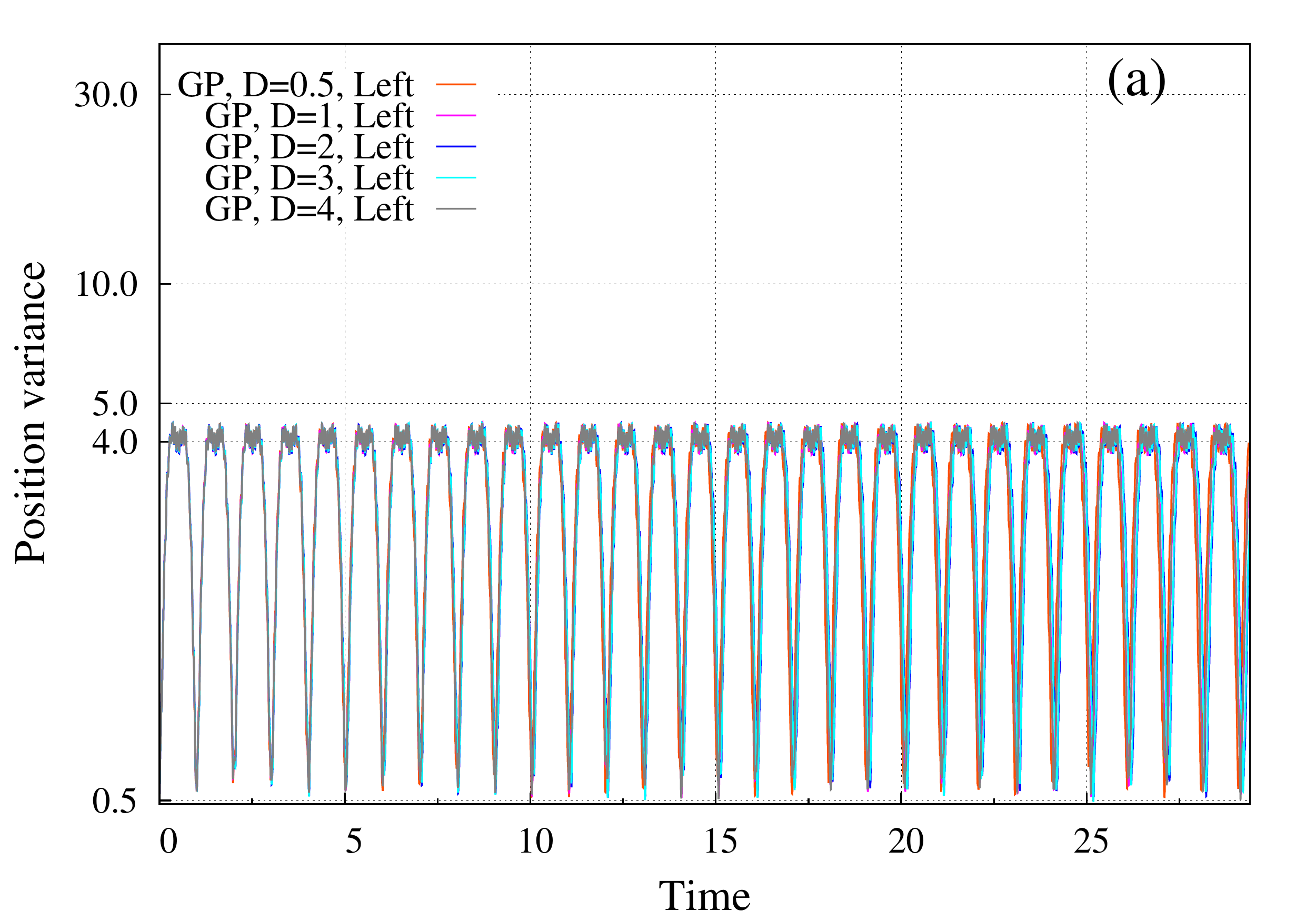} 
\includegraphics[width=0.32\linewidth, angle=270]{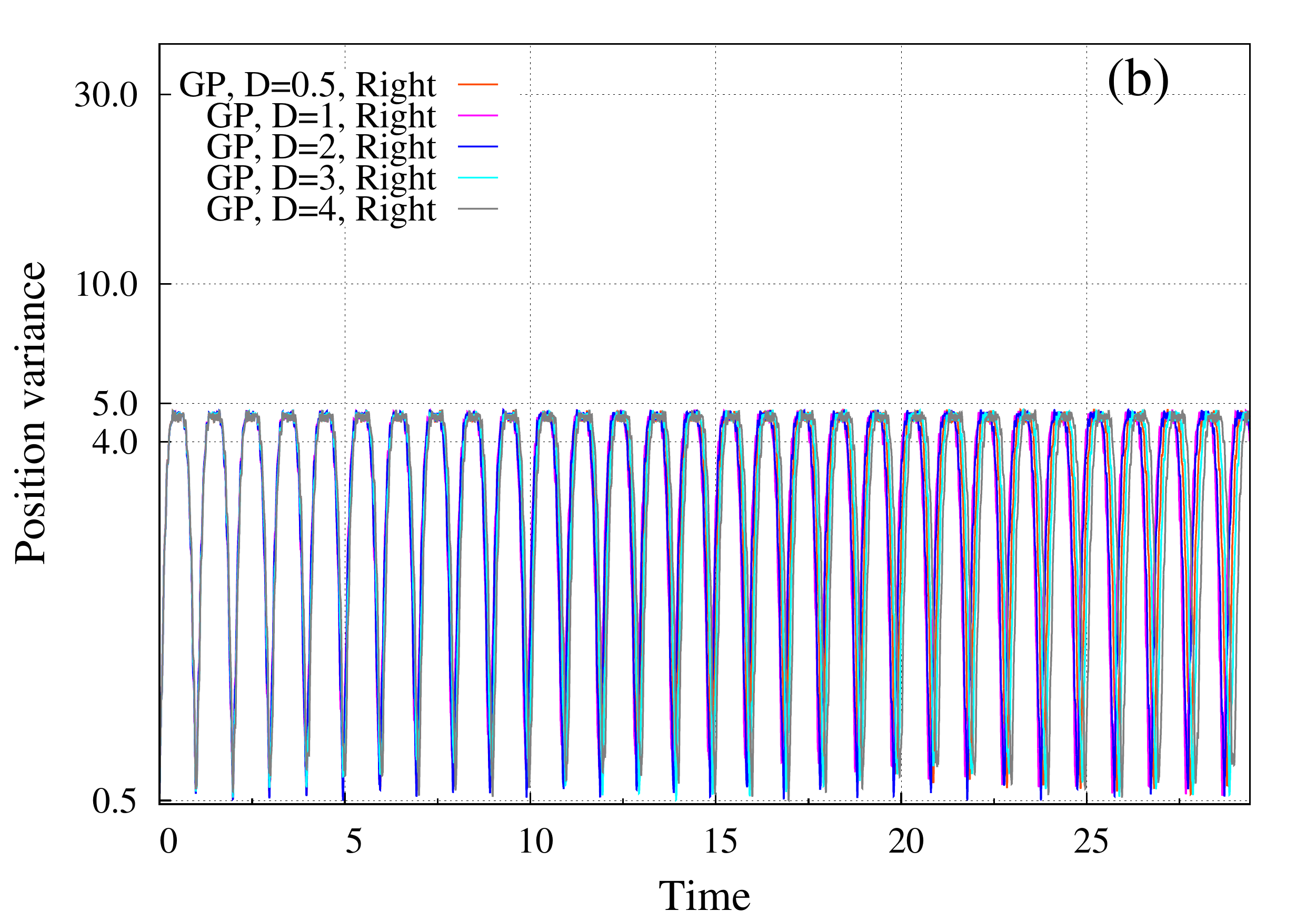} 
\end{center}
\caption{The many-body position variances $\frac{1}{N}\Delta_X^2(t)$ for different ranges $D$ of the interaction at the mean-field level. 
(a) For starting the dynamics from the left well and (b) for starting the dynamics from the right well. Time is scaled by $t_{Rabi}$. 
All quantities are dimensionless. Color codes are explained in each panel.}
\label{fig1}
\end{figure}

We plot the time evolution of $\frac{1}{N}\Delta_X^2(t)$ for different ranges of interaction for the left and the right well in Fig~\ref{fig1}(a) and (b),
respectively. We observe that for both wells, $\frac{1}{N}\Delta_X^2(t)$ oscillates smoothly with a frequency equal to the Rabi frequency. 
This is in contrast to the symmetric double well where the frequency of the oscillations in $\frac{1}{N}\Delta_X^2(t)$ is twice the Rabi cycle. 
Actually, unlike in the symmetric double
well, the density never completely tunnels out of the initial well of the asymmetric double well trap. This is manifested in the broadening of the peaks of oscillations
in $\frac{1}{N}\Delta_X^2(t)$. Also, the minima of the oscillations, after starting from $0.5$ is always slightly higher than $0.5$. Further, the curves for different
ranges $D$ of the interaction overlap with each other for both the wells, {though a very small deviation develops for the right well at a longer time. 
This shows that there is no significant effect of the ranges of interaction at the mean-field level.} 

\begin{figure}[!ht]
\begin{center}
\includegraphics[width=0.32\linewidth, angle=270]{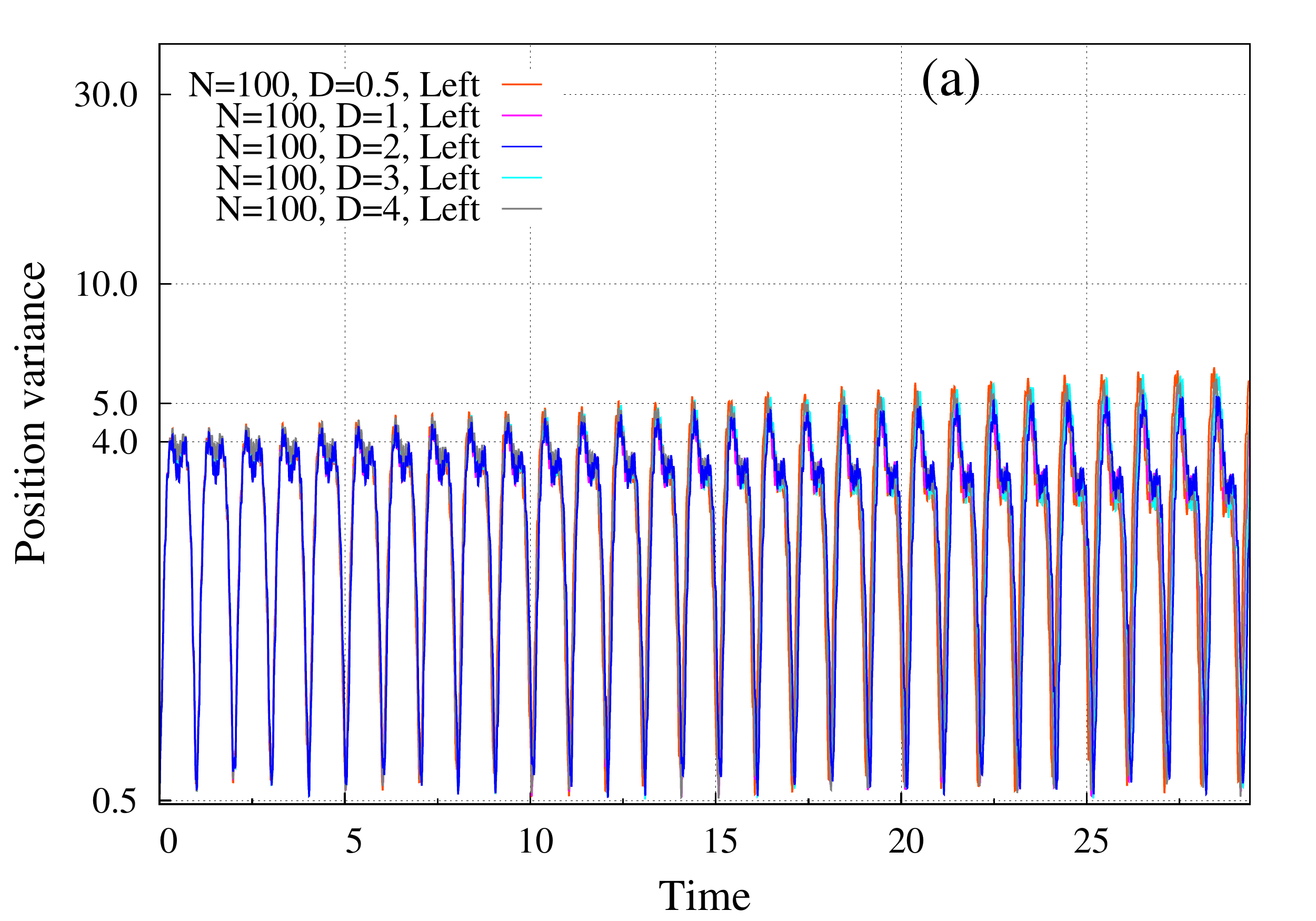} 
\includegraphics[width=0.32\linewidth, angle=270]{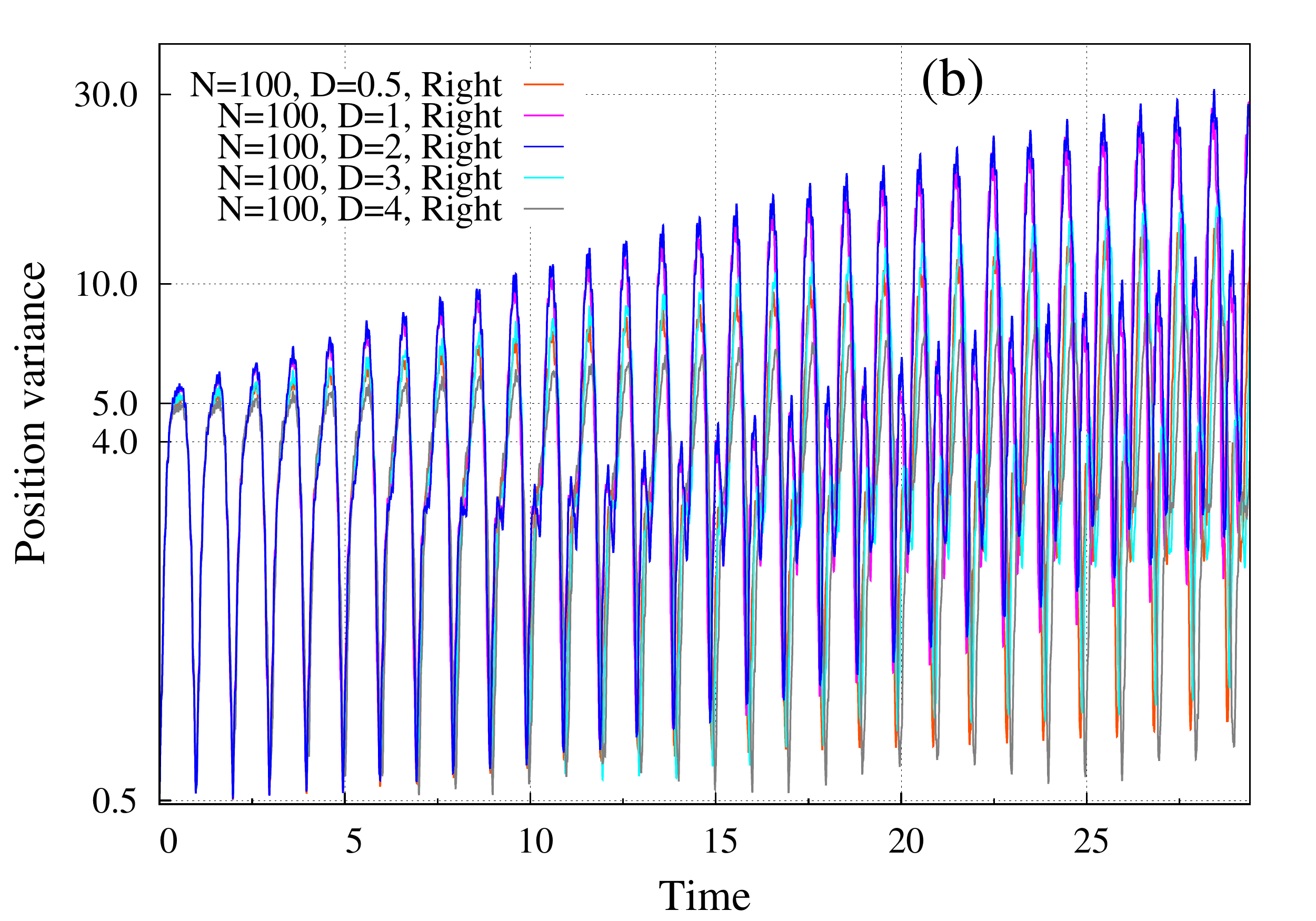}
\includegraphics[width=0.32\linewidth, angle=270]{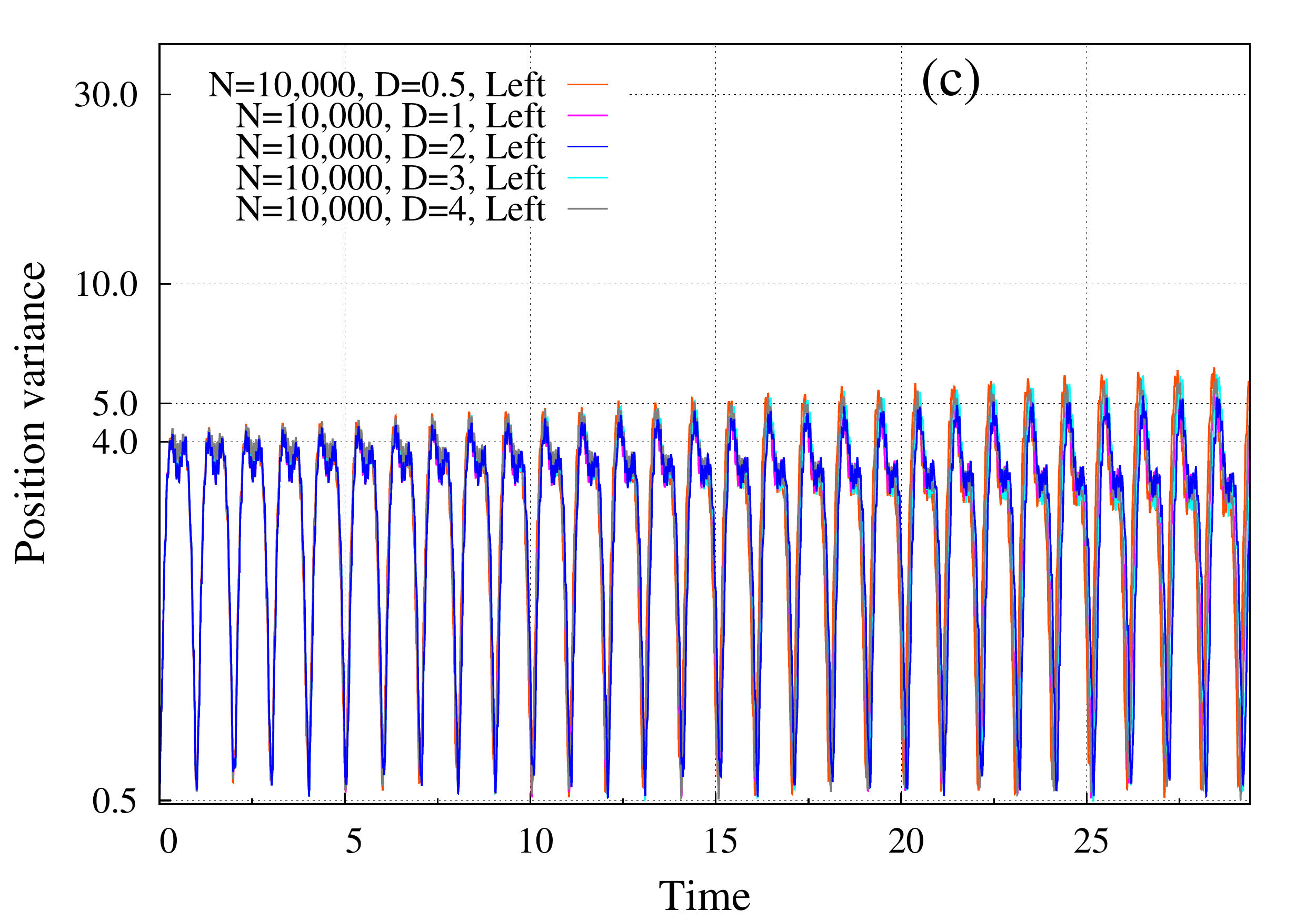} 
\includegraphics[width=0.32\linewidth, angle=270]{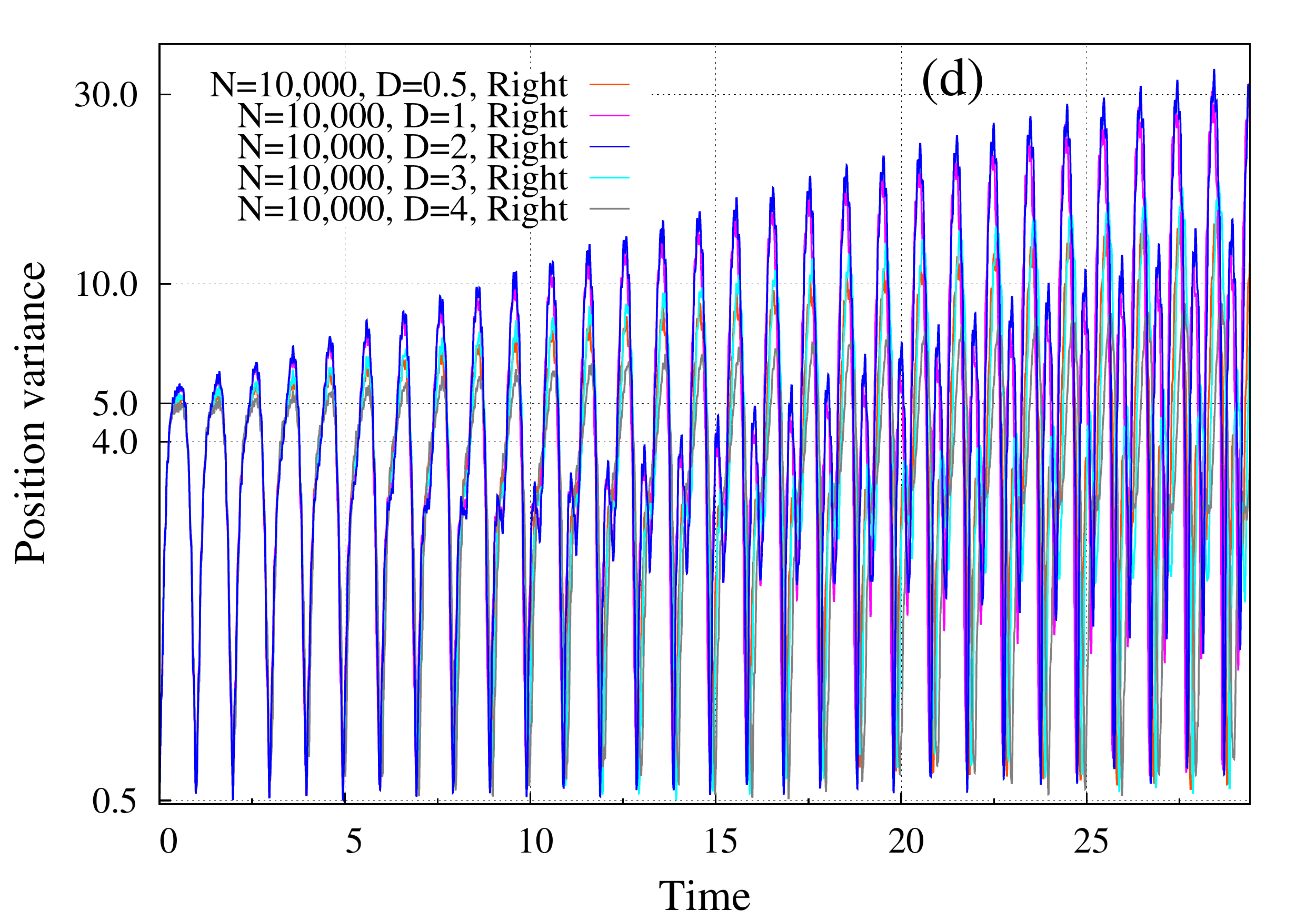}
\end{center}
\caption{The many-body position variance $\frac{1}{N}\Delta_X^2(t)$ obtained by the MCTDHB method with $M=2$ orbitals for different ranges $D$ of the interaction. 
(a) For a system of $N=100$ interacting bosons when the dynamics is started from the left well. (b) Corresponding dynamics when the initial condensate is 
prepared in the right well. (c) and (d) Corresponding time evolutions of $\frac{1}{N}\Delta_X^2(t)$ for a system of $N=10000$ interacting bosons when the 
initial state is prepared in the left and the right well respectively. Time is scaled by $t_{Rabi}$. All quantities are dimensionless. Color codes are explained in each panel. }
\label{fig2}
\end{figure}

Next, we study the time evolution of the $\frac{1}{N}\Delta_X^2(t)$ at the many-body level. We computed the many-body results by the MCTDHB method
with $M=2$ orbitals, see \ref{MCTDHB} for details. In Fig.~\ref{fig2}(a) and (b) we have plotted the time evolution of the many-body position variance
of a system of $N=100$ interacting bosons for various ranges of the interaction, for the left well and the right well, respectively. We observe that 
$\frac{1}{N}\Delta_X^2$ oscillates with the Rabi frequency between a growing maximum and a minimum which is approximately equal to $0.5$.
The deviation from $0.5$ is because of the remnants in each well as discussed above. This is further manifested in the splitting of the peaks into
two sub peaks. Moreover, the difference between the two sub peaks grows with time for both wells. We also see a clear impact of the range $D$ of the 
interaction on the time evolution of $\frac{1}{N}\Delta_X^2(t)$  though $D$ affects the the time evolution of the many-body position variance 
$\frac{1}{N}\Delta_X^2(t)$ differently for the left $V_L(x)$ and the right $V_R(x)$ wells. While for $V_L(x)$, the peak values of the oscillations 
in $\frac{1}{N}\Delta_X^2(t)$ decrease with increasing $D$ until $D=l/2$ for which the peak values are minimum. On the other hand, similarly to 
the symmetric double well, the peak values of $\frac{1}{N}\Delta_X^2(t)$ increase with increasing $D$ until $D=l/2$ for which it reaches the maximum. 
However, here, the minima also increases with time for all $D$. Further, while the higher sub peak appears on the left side for $V_L(x)$, it is on the right side for $V_R(x)$.  

\begin{figure}[!ht]
\begin{center}
\includegraphics[width=0.32\linewidth, angle=270]{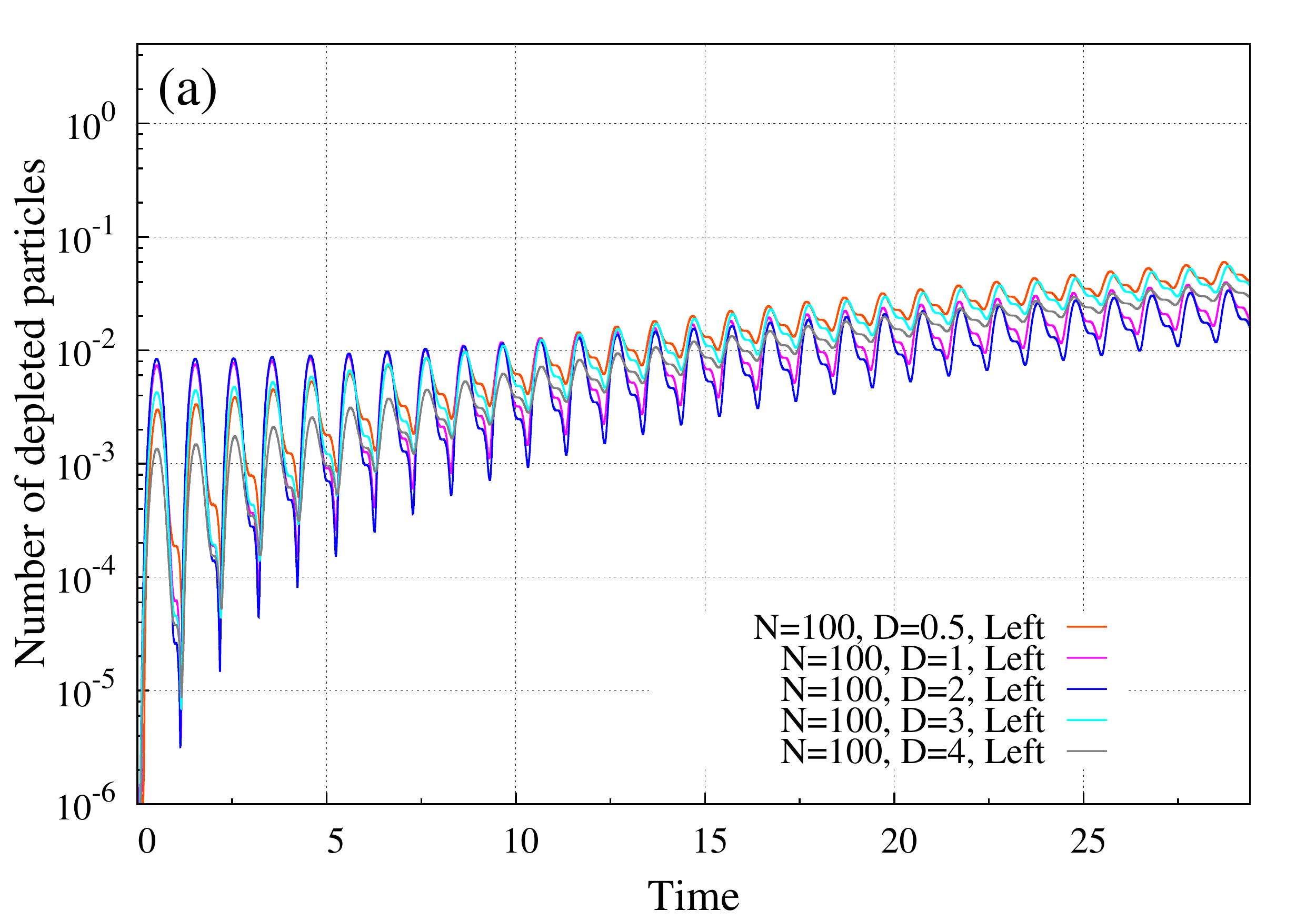} 
\includegraphics[width=0.32\linewidth, angle=270]{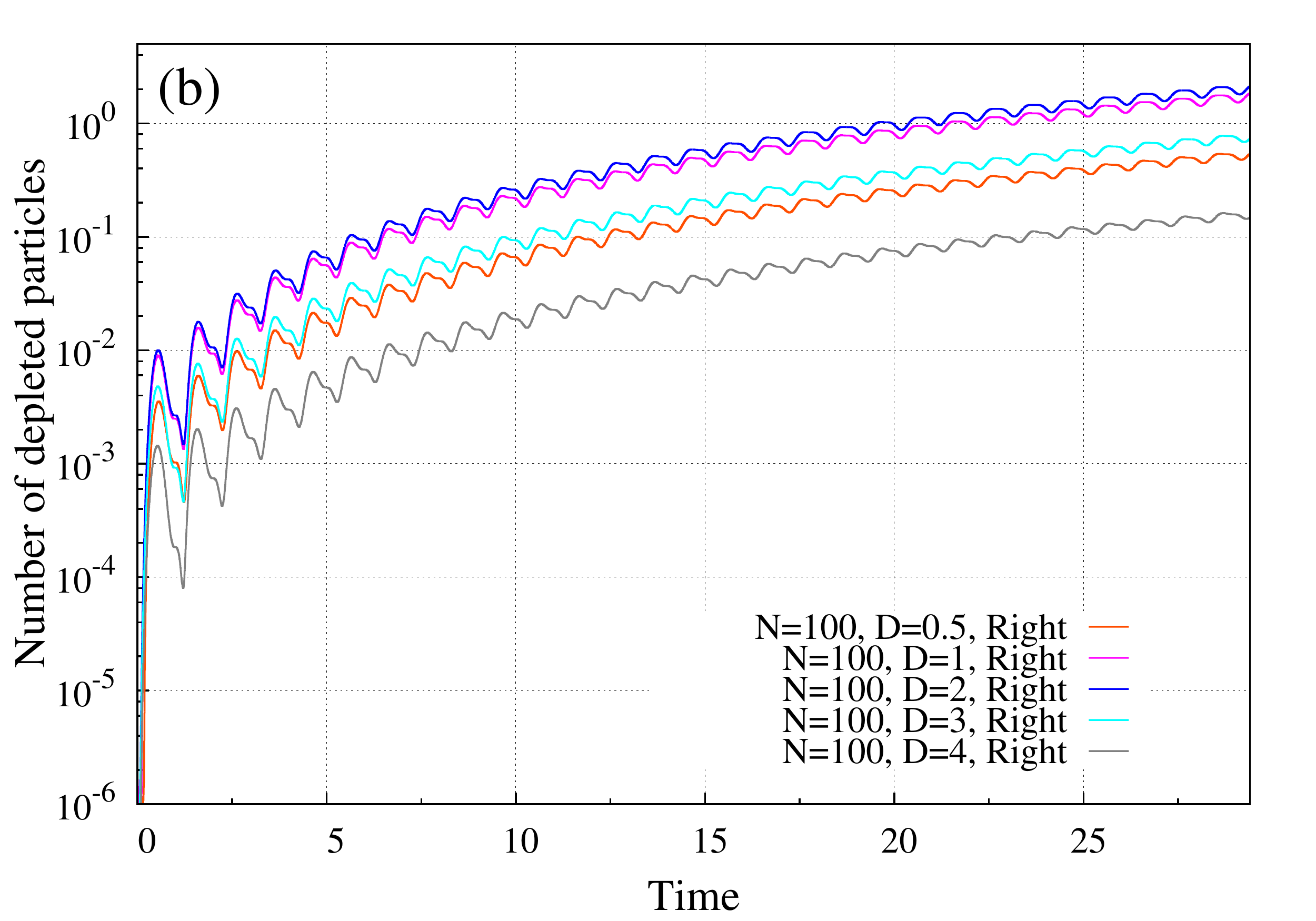} 
\includegraphics[width=0.32\linewidth, angle=270]{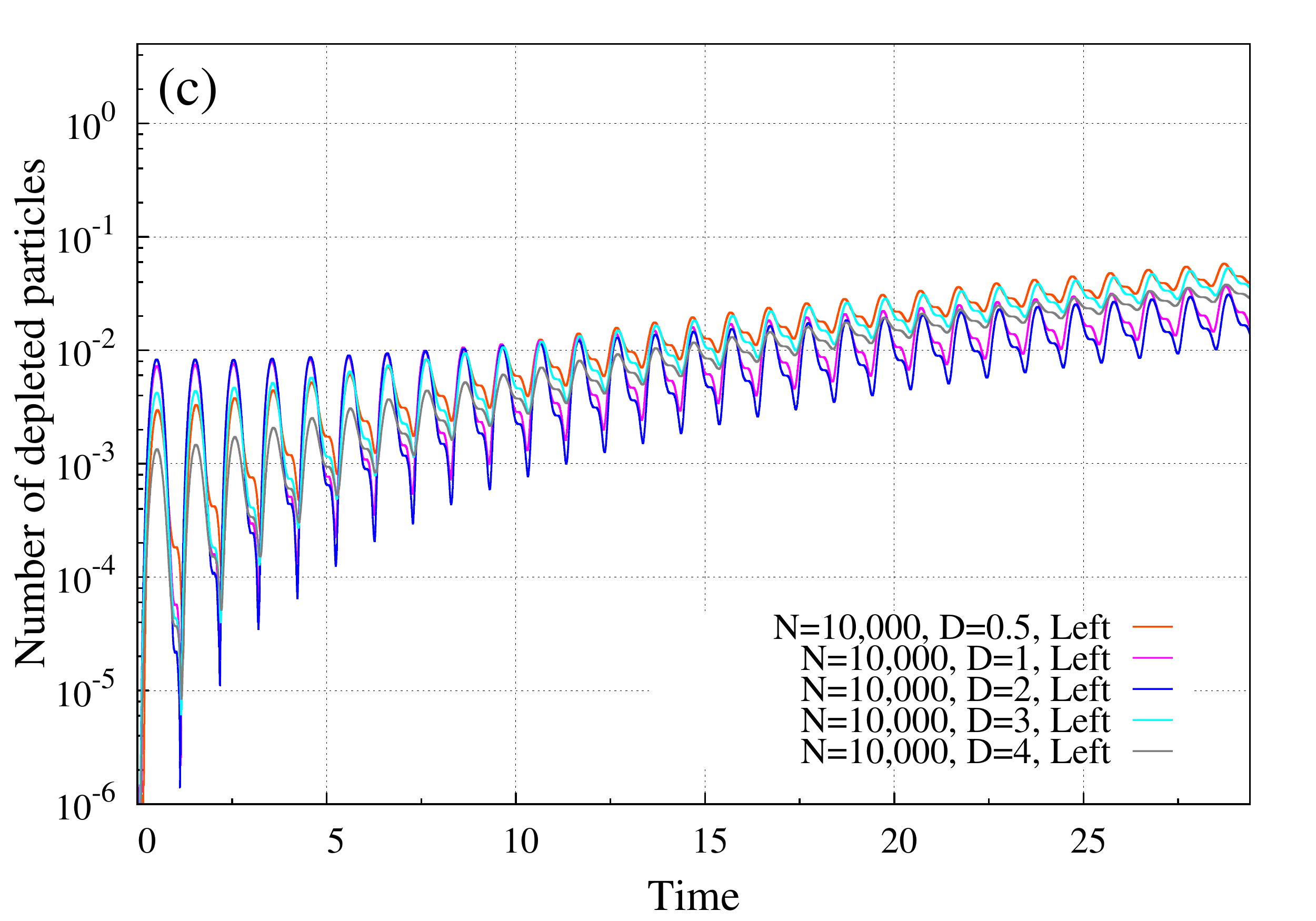} 
\includegraphics[width=0.32\linewidth, angle=270]{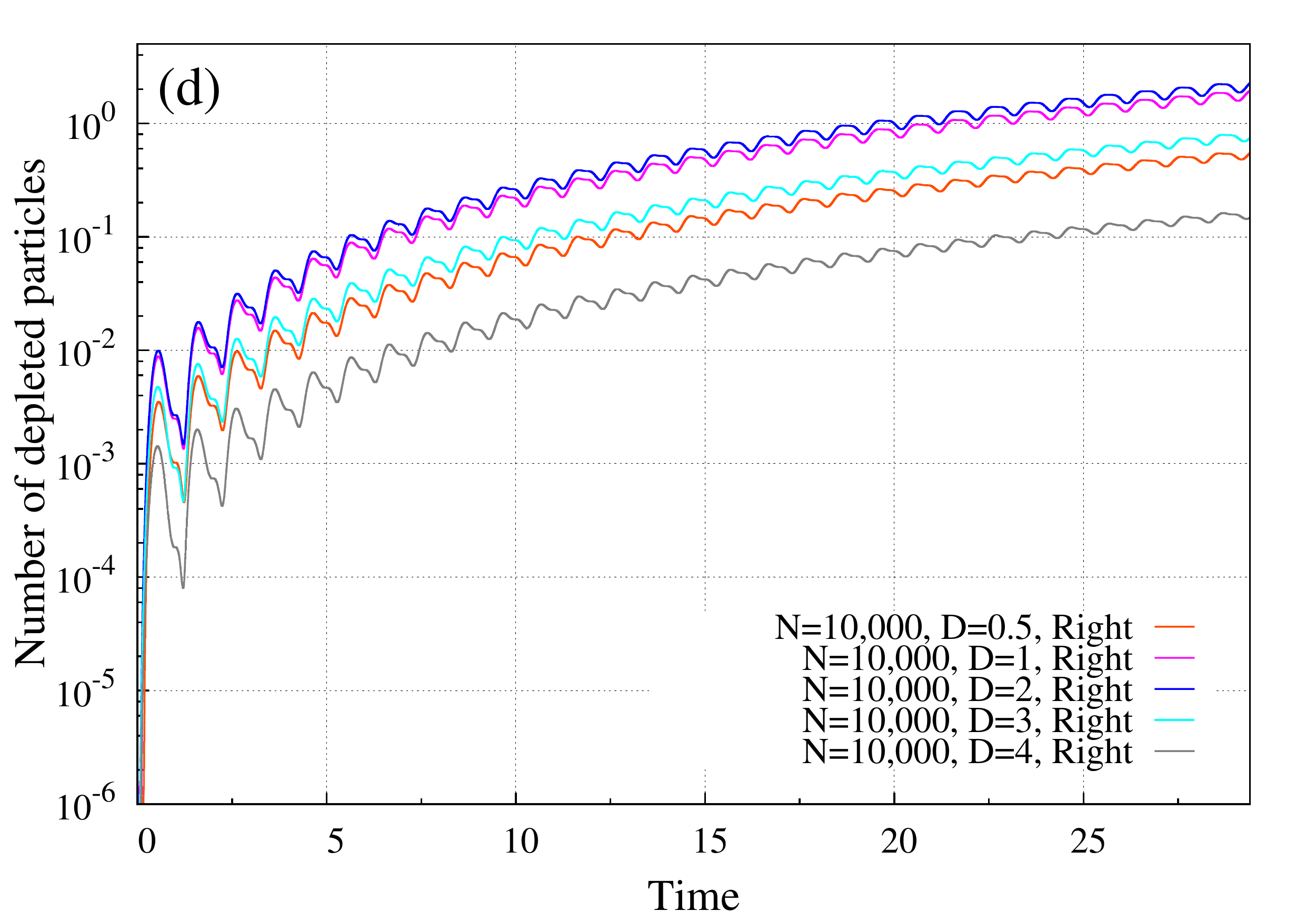} 
\end{center}
\caption{The number of depleted atoms outside the condensate mode for different ranges $D$ of the interaction. The data have been obtained by the MCTDHB with $M=2$ orbitals. (a) and (b) For a system of $N=100$ interacting bosons for the left and the right wells respectively. 
Corresponding number of depleted atoms for a system of $N=10000$ bosons when the dynamics is started from the left [panel (c)] and the right 
[panel (d)] well. Time is scaled by $t_{Rabi}$. The quantities shown here are dimensionless. Color codes are explained in each panel.}
\label{fig3}
\end{figure}

Now, to confirm that at this interaction strength for $N=100$, the system is essentially condensed, we have also plotted the number of depleted 
atoms outside the condensed mode as a function of time for the right as well as the left wells in Fig.~\ref{fig3}(a) and (b), respectively. {We observe that
the number of depleted atoms grows with time in an oscillatory manner}. We also note that the growth rates of 
the number of depleted atoms are different for $V_L(x)$ and $V_R(x)$. {Moreover, for $V_L(x)$ itself,} these growth rates are different 
depending on the range $D$ of the interaction. Accordingly, at short time, the system is the most depleted for $D=l/2$ whereas at 
longer time, the depletion is the least for $D=l/2$. {Also for the left well, the depletions of the system for all $D$ are approximately of the same order,
though they vary a bit depending on $D$}. On the other hand, for $V_R(x)$ the growth rate of the depletion  is independent of $D$. However, the number of 
depleted particles varies widely for different $D$.  Also, at any point of time, the number of depleted particles increases with $D$ and is maximum for $D=l/2$. 
With further increase in $D$, the number of depleted particles decreases. Further, for all $D$, the system is more depleted for $V_R(x)$ than $V_L(x)$. 
Finally, for all cases, the number of depleted particles are negligibly small implying that the system is essentially condensed and the mean-field GP theory 
should be adequate to describe the dynamics of the system. {Thus, the effects of the range $D$ of the interaction observed above in the time evolution of
$\frac{1}{N}\Delta_X^2$ 
appear only at the many-body level, even when the mean-field theory should be applicable for the description of the out-of-equilibrium dynamics of the system's density per particle}.  

Rigorous results exist in the literature proving that, in the infinite-particle limit keeping $\Lambda$ fixed, the density per particle and the energy per particle of
the ground state of the system coincide with the corresponding mean-field GP results~[27-30]. 
Therefore, next we examine if the effects of the ranges of interaction
and the asymmetry of the trap, observed at the many-body level for a finite size system, still persist for a much larger system size {\textit{en route} to 
the infinite-particle
limit}. Hence, next we consider a system of $N=10000$ bosons keeping $\Lambda=0.01$ fixed. We again study the time evolution of the many-body position variance of the 
system by MCTDHB method with $M=2$ orbitals. The $\frac{1}{N}\Delta_X^2(t)$ for various ranges of interaction $D$ for both $V_L(x)$ and $V_R(x)$ are shown 
in Fig.~\ref{fig2}(c) and (d), respectively. We observe essentially identical behavior in the oscillatory growth of $\frac{1}{N}\Delta_X^2(t)$ with time for $V_L(x)$ as 
for the system with $N=100$ particles. However for $V_R(x)$, though $\frac{1}{N}\Delta_X^2(t)$ exhibit an oscillatory growth similarly to that of a system 
with $N=100$ bosons, here the minima of $\frac{1}{N}\Delta_X^2(t)$ do not grow with time and is only slightly higher than $0.5$. 
We also note that for the right well, the time evolution of $\frac{1}{N}\Delta_X^2(t)$ for all $D$ are qualitatively  similar to the corresponding symmetric 
double well case except that the growth rate, frequencies, and amplitudes are different.   

Since, as already mentioned, the many-body position variance depends on the actual number of depleted atoms, next we also plot the corresponding number of 
depleted particles for $N=10000$ as a function of time for both the left and the right wells in Fig.~\ref{fig3}(c) and (d), respectively. We find that the oscillatory 
growth of the number of depleted particles are practically identical for both $N=100$ and $N=10000$ bosons. Further, at any point of time, the corresponding numbers of depleted 
atoms for all $D$ and wells are also essentially identical for both $N=100$ and $N=10000$. This implies that at any point of time the system with $N=10000$ is actually two 
orders of magnitudes less depleted and therefore closer to the mean-field approximation. Thus the practically identical behaviors of $\frac{1}{N}\Delta_X^2(t)$ for $N=100$ and 
$N=10000$ for all cases indicate that the effects of the range $D$ of interaction have converged with respect to $N$ and will, therefore, persist in the infinite-particle 
limit also. This implies that the effects of the range $D$ of the interaction on the asymmetric BJJ observed here are many-body in nature and can 
not be described by the mean-field theory even when the system is $100\%$ condensed.

\section{Summary}
\label{Summery}
\hspace*{0.5cm}
In summary, we have studied the impact of a finite range interaction on the dynamics of an interacting Bose gas in a one dimensional asymmetric double well trap. 
We have characterized the dynamics by the time evolution of the many-body position variance $\frac{1}{N}\Delta_X^2(t)$ which can deviate from the corresponding 
mean-field results even when one out of a million atoms is outside the condensed mode. We focused only on the weakly interacting regime where the system is 
essentially fully condensed and the mean-field theory should be adequate for the description of its the out-of-equilibrium dynamics. 
Also, we examined the infinite-particle limit where the system is $100\%$ condensed and its ground state energy per particle and the ground state density per 
particle coincide with the corresponding mean-field GP results. We find that, even for a weakly interacting system, the effect of the range of interaction on 
the dynamics of a one dimensional asymmetric BJJ can be described only at the many-body level. {Further, such many-body effects are also found to 
persist in the large particle systems when the infinite-particle limit is essentially achieved.} Also, the range of the interaction affects the dynamics 
differently depending on whether 
the initial condensate state is prepared in the left well or in the right well. While the dynamics for the right well is qualitatively similar to the symmetric 
BJJ dynamics, differences appear in the growth rate, frequencies, and peak values of $\frac{1}{N}\Delta_X^2(t)$. On the other hand, completely new features are 
observed for the left well.

\ack
\hspace*{0.5cm}
This research was supported by the Israel Science Foundation (Grant No. 600/15). Computation
time on the High Performance Computing system Hive of the Faculty of Natural
Sciences at University of Haifa is gratefully acknowledged. We also acknowledge computation time on the BwForCluster and 
the Cray XC40 system Hazelhen at the High Performance
Computing Center Stuttgart (HLRS). SKH gratefully
acknowledges the continuous hospitality at the Lewiner Institute for Theoretical Physics
(LITP), Department of Physics, Technion - Israel Institute of Technology. SKH also acknowledges 
fruitful discussions with Anal Bhowmik. 

\appendix
\section{Multiconfigurational time-dependent Hartree method for bosons (MCTDHB)}\label{MCTDHB}
\hspace*{0.5cm}

In this work, we have used the multiconfigurational time-dependent Hartree method for bosons (MCTDHB) for studying the dynamics of an interacting Bose gas in a one dimensional
asymmetric double well. MCTDHB is an in-principle numerically exact method which has been developed~\cite{Ofir2008, Streltsov2007} for solving the 
time-dependent many-body Schr\"odinger equation and benchmarked with an exactly-solvable model~\cite{Lode2012,Axel_MCTDHF_HIM}. 
This method has already been extensively used in the literature~[13,34-42]. 

MCTDHB is a time-dependent variational method in which the ansatz is chosen as the superposition of all possible configurations obtained by distributing
$N$ bosons in $M$ time-dependent single-particle orbitals, viz.,
\beq\label{MCTDHB_Psi}
\left|\Psi(t)\right> = 
\sum_{\vec{n}}C_{\vec{n}}(t)\left|\vec{n};t\right>.
\eeq
Here the occupations $\vec{n}=(n_1,n_2,\ldots,n_M)$ preserve the total number of bosons $N$.
With this ansatz, the time-dependent action is minimized to obtain the working equations for the orbitals and the
coefficients. For details, we refer to Ref.~\cite{Ofir2008}. Here we point out that for $M=1$ orbital, we get the ansatz for the GP equation while in the 
limit $M \rightarrow \infty$ the ansatz expands over the full Hilbert space and hence gives an exact theory. However, in practice, one has to use a finite number $M$ of orbitals. 
In our numerical calculations, we keep on repeating the computations with increasing $M$ until the convergence is reached and thereby we obtain the numerically highly accurate results, see~\ref{convergence}. 
We point out that because of the time-dependent permanents,  one can use a much shorter expansion in Eq.~(\ref{MCTDHB_Psi}) than if only the coefficients are allowed 
to be time-dependent, and this leads to a significant computational advantage. 
The ground state properties of the system can be obtained by propagating the MCTDHB equations in imaginary time~\cite{MCHB}.

\section{Details of the numerical computations and their convergence}\label{convergence}
\hspace*{0.5cm}
For our numerical computations, we have used 
the numerical implementation of the parallel version of MCTDHB~\cite{Streltsov1, Streltsov2}.
Here we have represented the many-body Hamiltonian by 128 exponential 
discrete-variable-representation (DVR) grid points using a fast Fourier transformation (FFT) routine within a box of size [-10:10). We obtain the initial many-body ground state of the
BEC in both the left and right wells by propagating the MCTDHB equations in imaginary time. 
As already mentioned above, in our numerical calculations we keep on increasing the number of time-adaptive orbitals $M$ to obtain numerically exact results. Below, we 
explicitly show the numerical convergence of our results with respect to the number of orbitals $M$ for a system of $N=10$ bosons and $\Lambda=0.01$.
Since increasing $N$, keeping $\Lambda$ fixed, amounts
to an effectively weaker interaction, the convergences of our results for the systems of $N=100$ and $N=10000$ bosons 
are ensured by this procedure.

\begin{figure}[!ht]
	\begin{center}
			\includegraphics[width=0.32\linewidth, angle=270]{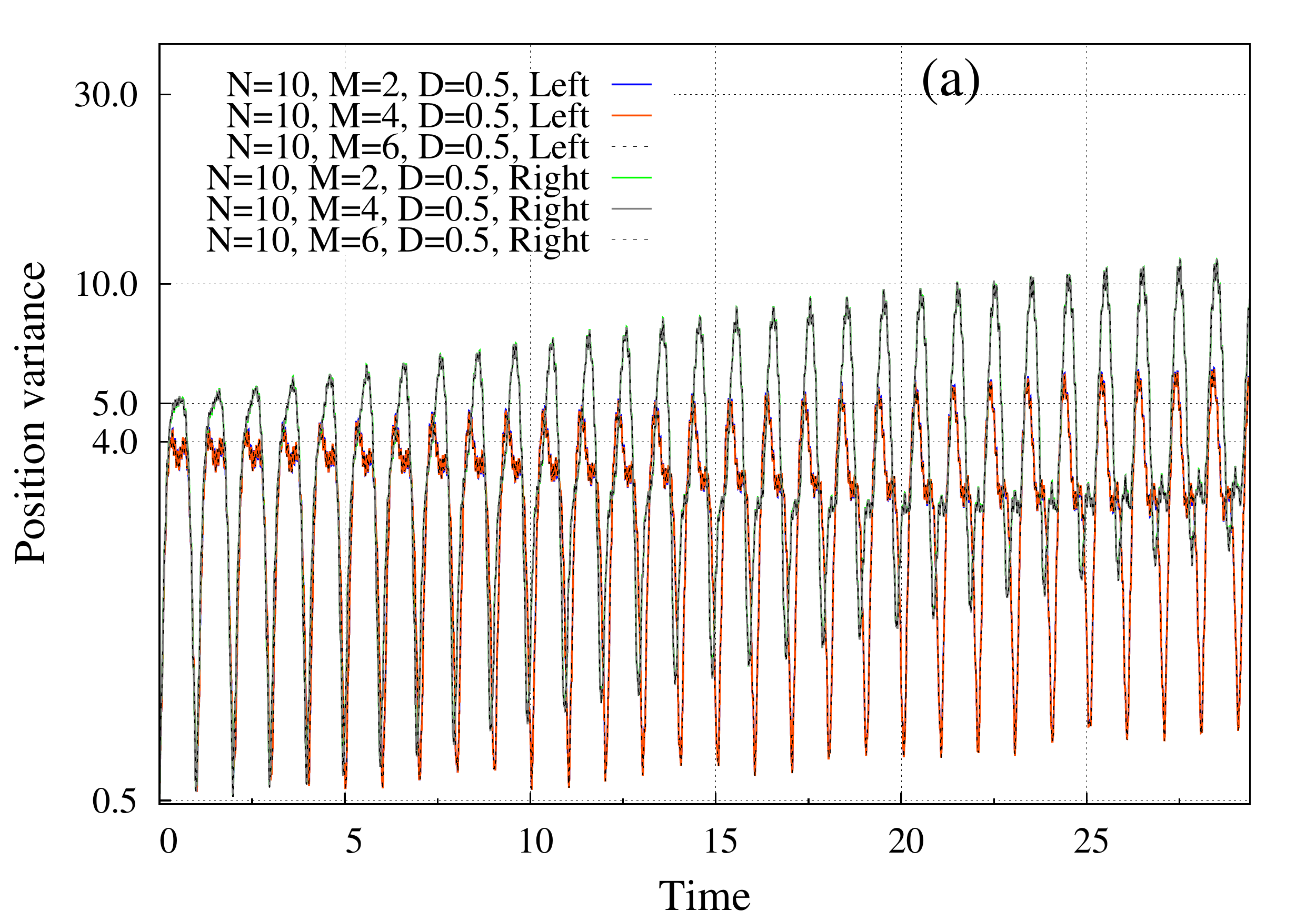} 
			\includegraphics[width=0.32\linewidth, angle=270]{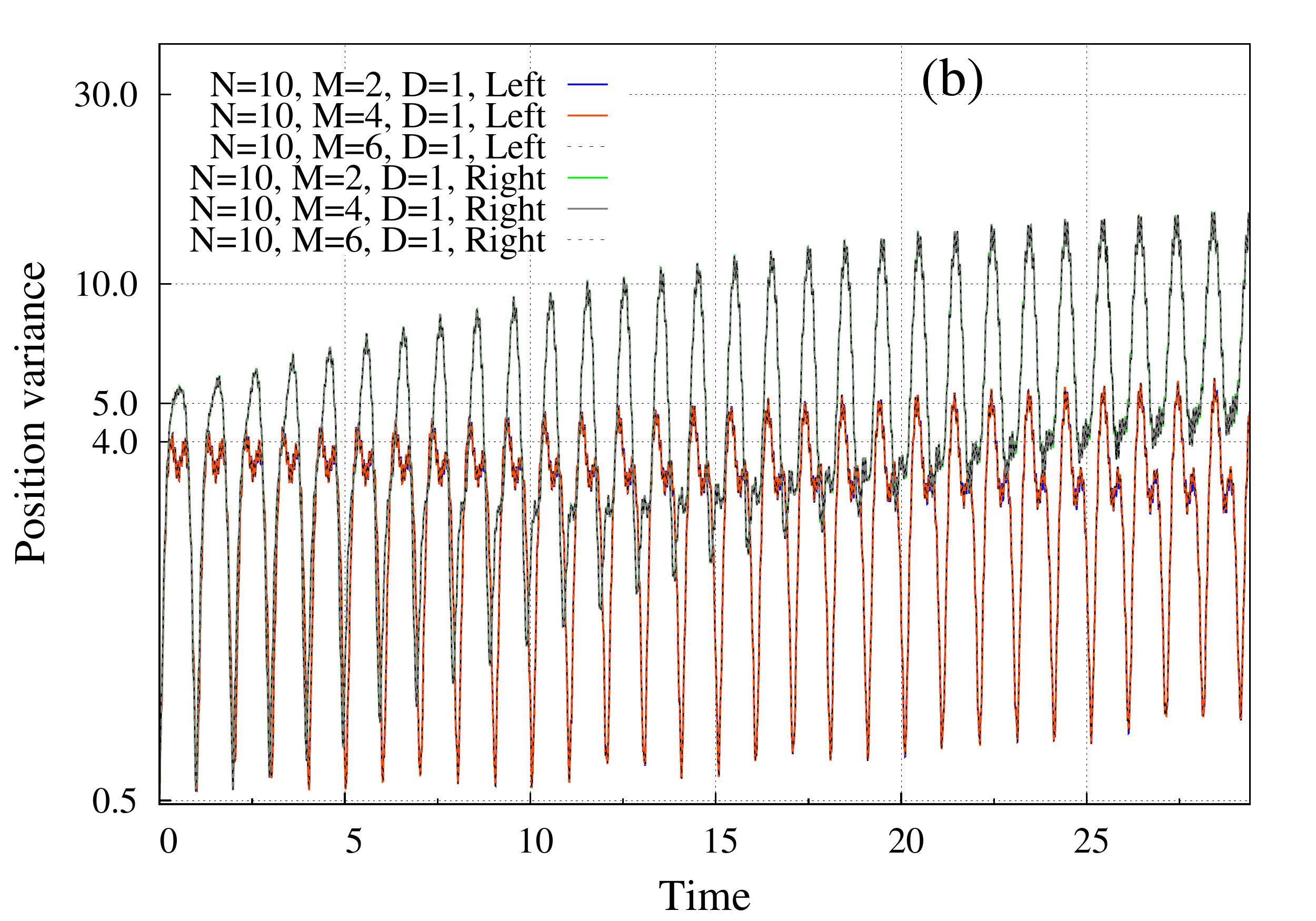} 
			\includegraphics[width=0.32\linewidth, angle=270]{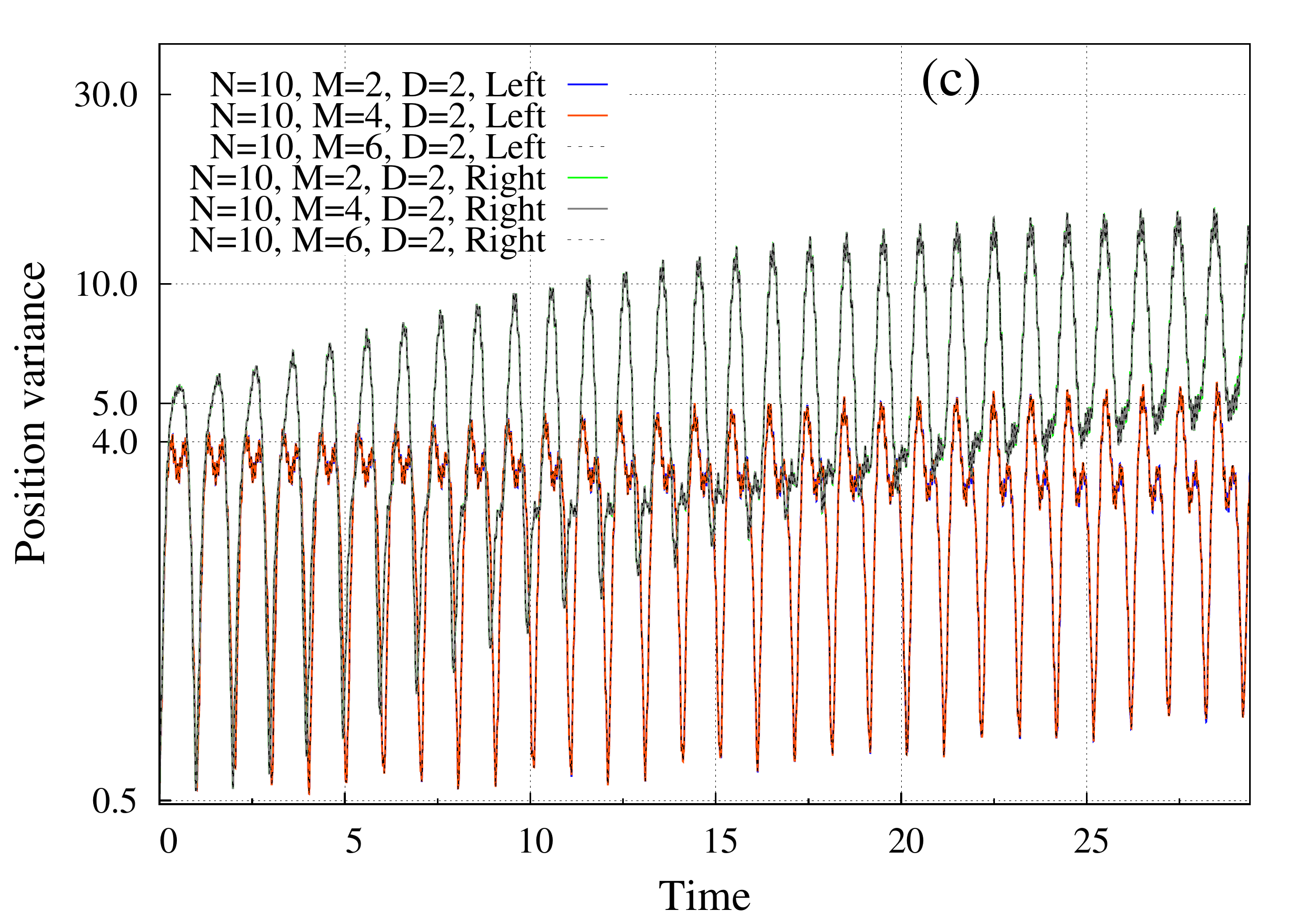} 
			\includegraphics[width=0.32\linewidth, angle=270]{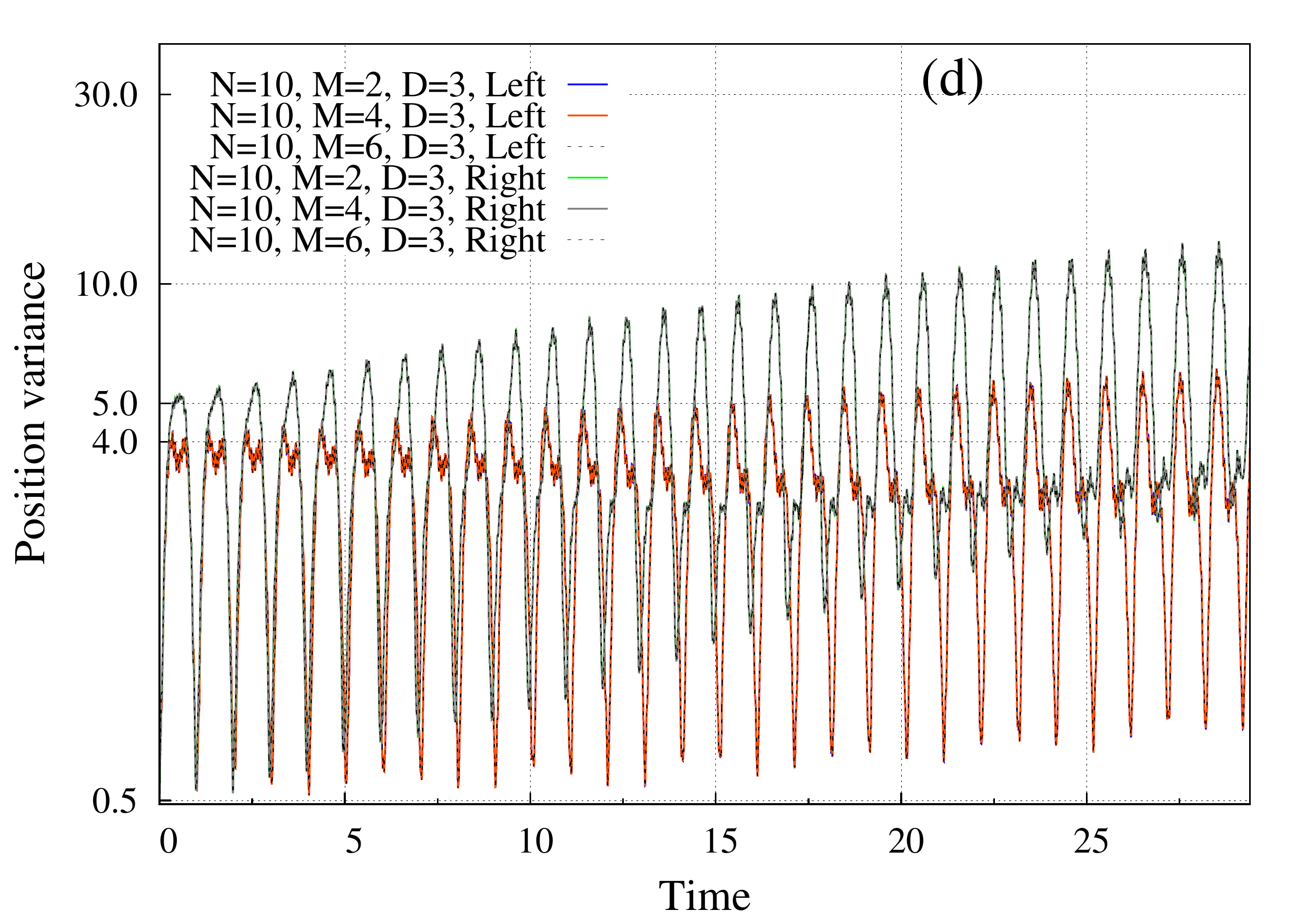} 
			\includegraphics[width=0.32\linewidth,angle=270]{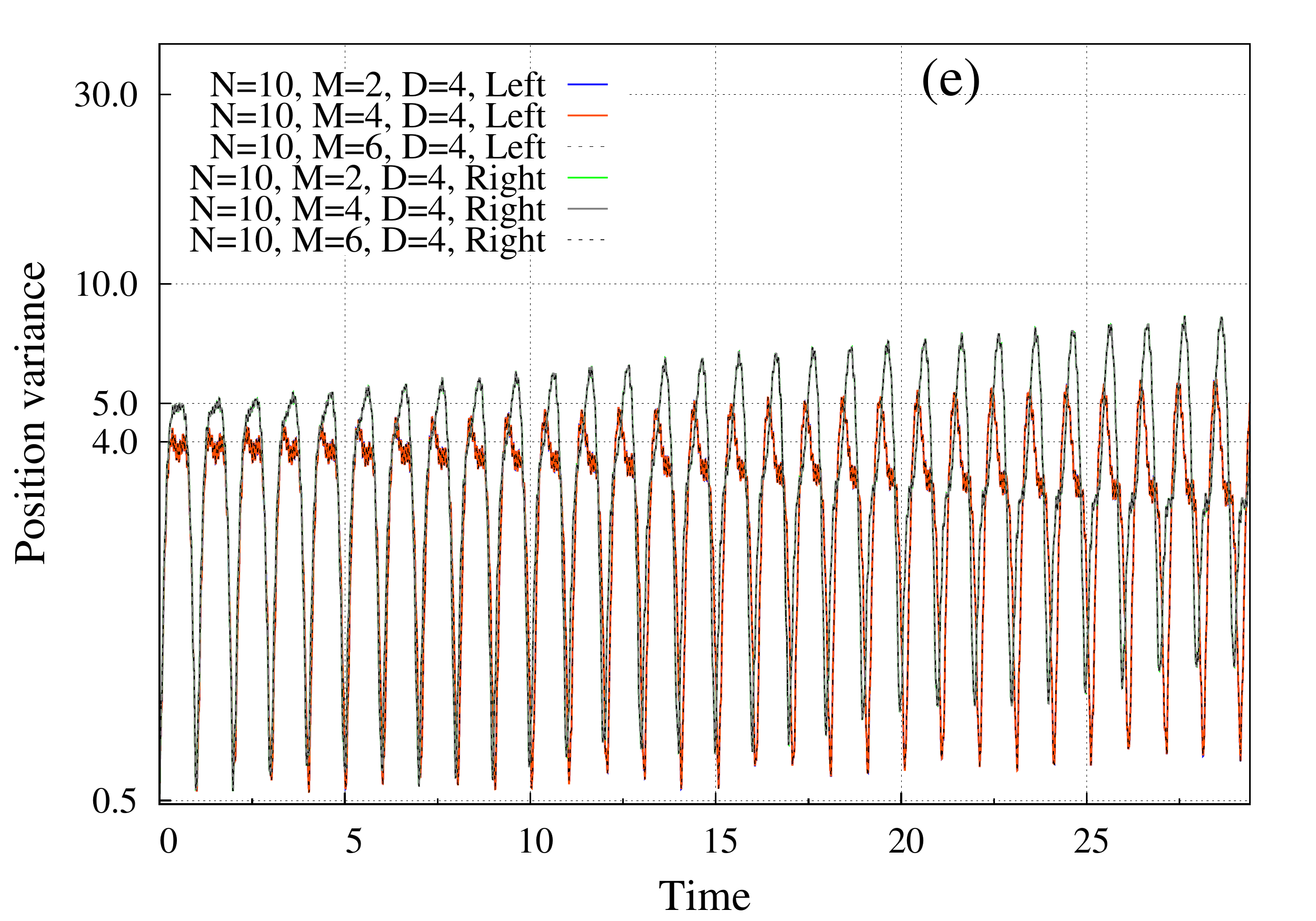} 
	\end{center}
	\caption{Convergence of $\frac{1}{N}\Delta_X^2(t)$ for both wells for (a) $D=l/8$, (b) $D=l/4$, (c) $D=l/2$, (d) $D=3l/4$ and (e) $D=l$. $l=4$ is the length scale of the system and time is scaled by $t_{Rabi}$. See text for further details. All quantities are dimensionless. Color codes are explained in each panel.}
	\label{fig4}
\end{figure}

We demonstrate the numerical convergence of $\frac{1}{N}\Delta_X^2(t)$ for each $D$ separately for both the wells. This also ensures that the small differences observed among 
$\frac{1}{N}\Delta_X^2(t)$ and the number of depleted atoms for different $D$ are well converged. 
In Fig.~\ref{fig4}, each panel shows the results for $\frac{1}{N}\Delta_X^2(t)$ for each $D$ obtained with $M=2,4,$ and $6$ orbitals for both wells. We find that for all cases, 
$\frac{1}{N}\Delta_X^2(t)$ exhibit an oscillatory growth with the frequency of oscillation being equal to the Rabi frequency. However, now the minima of $\frac{1}{N}\Delta_X^2(t)$
grow with time for the left well also in addition to that for the right well. Also, $\frac{1}{N}\Delta_X^2(t)$ computed with $M=2,4,$ and $6$ practically overlap with each other 
for all cases with the results for $M=4$ and $6$ lying atop each other. This shows that the use of $M=2$ orbitals already gives the accurate description of the time evolution of $\frac{1}{N}\Delta_X^2(t)$ 
for all cases.  

\begin{figure}[!ht]
	\begin{center}
			\includegraphics[width=0.32\linewidth, angle=270]{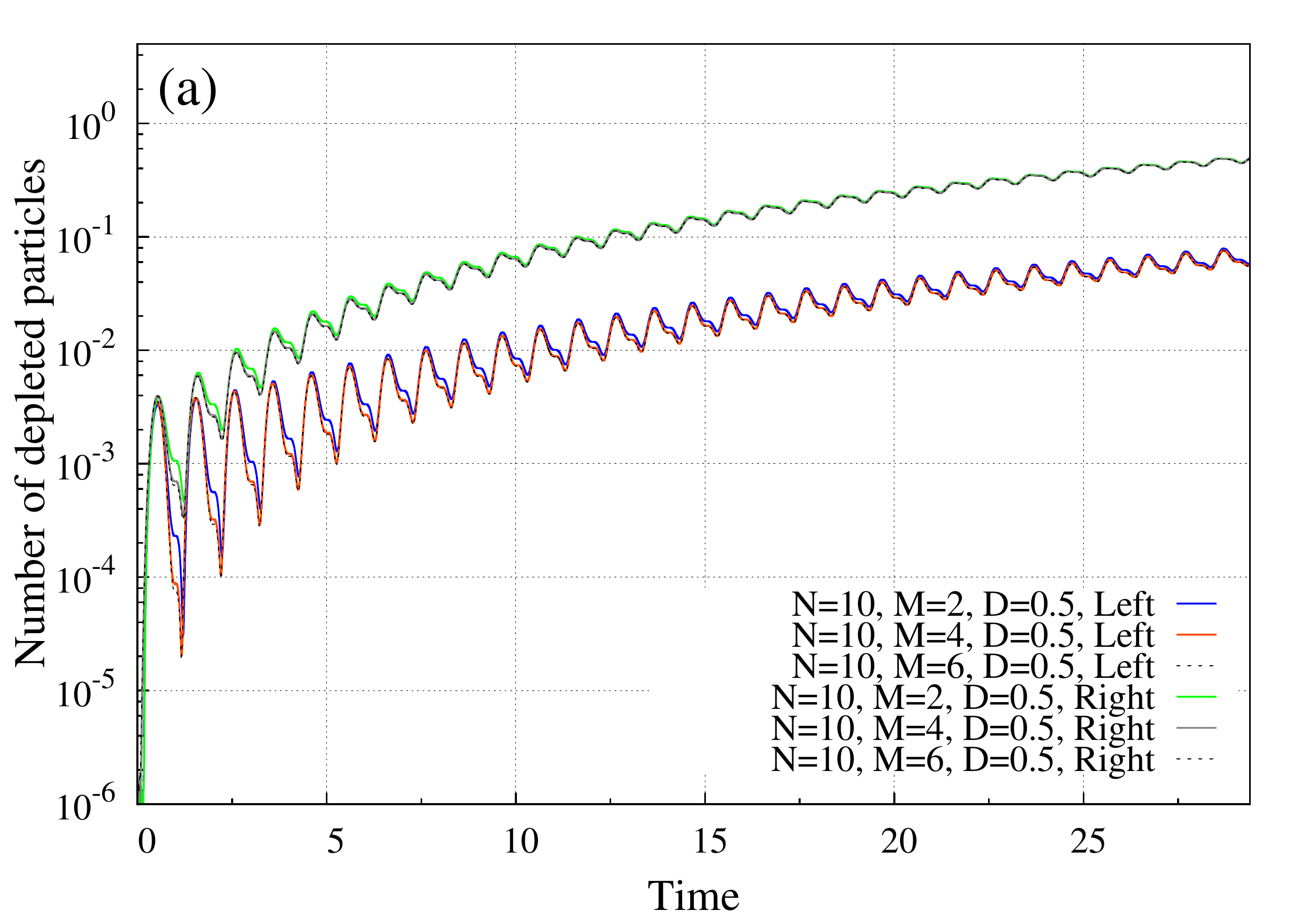} 
			\includegraphics[width=0.32\linewidth, angle=270]{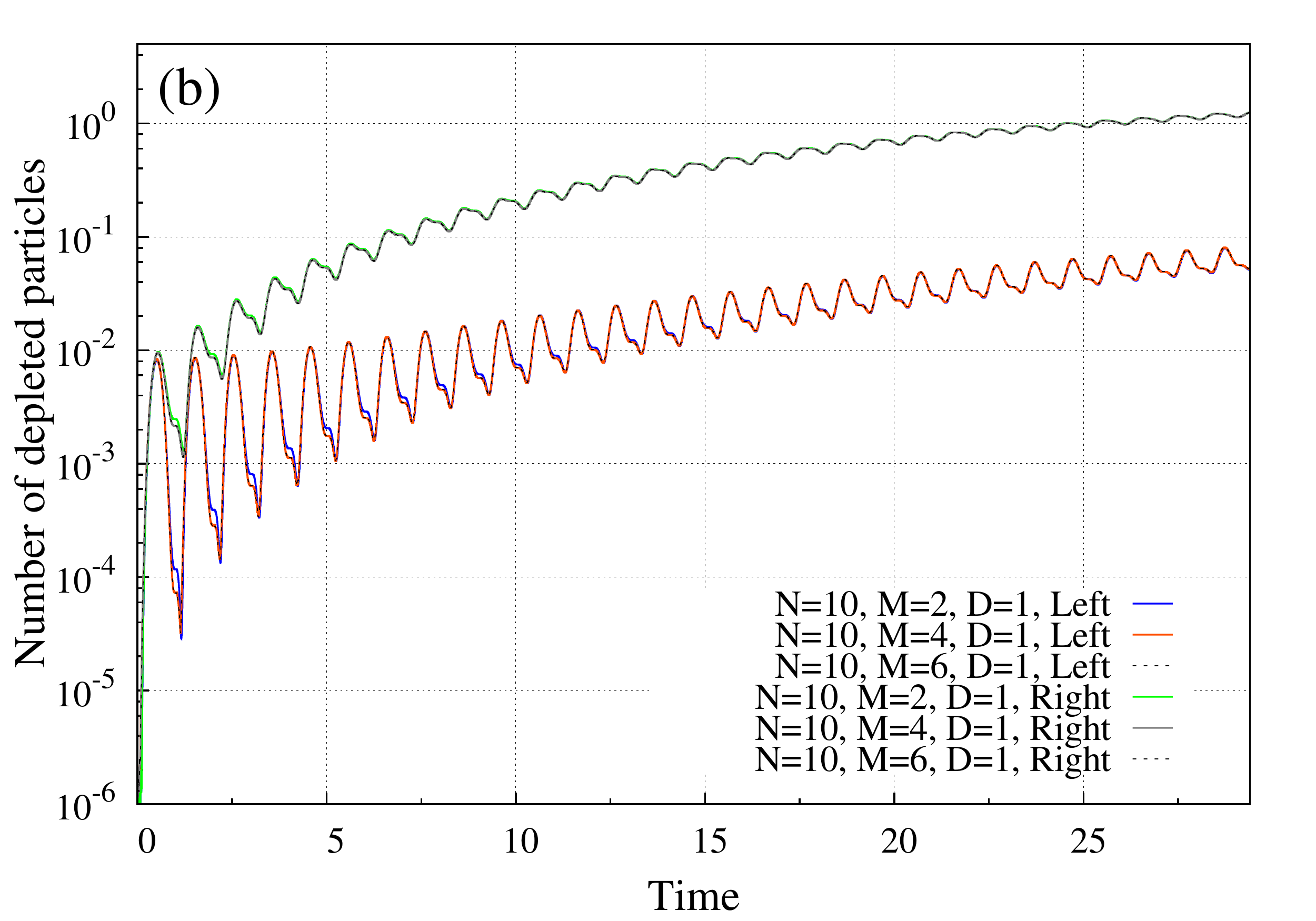} 
			\includegraphics[width=0.32\linewidth, angle=270]{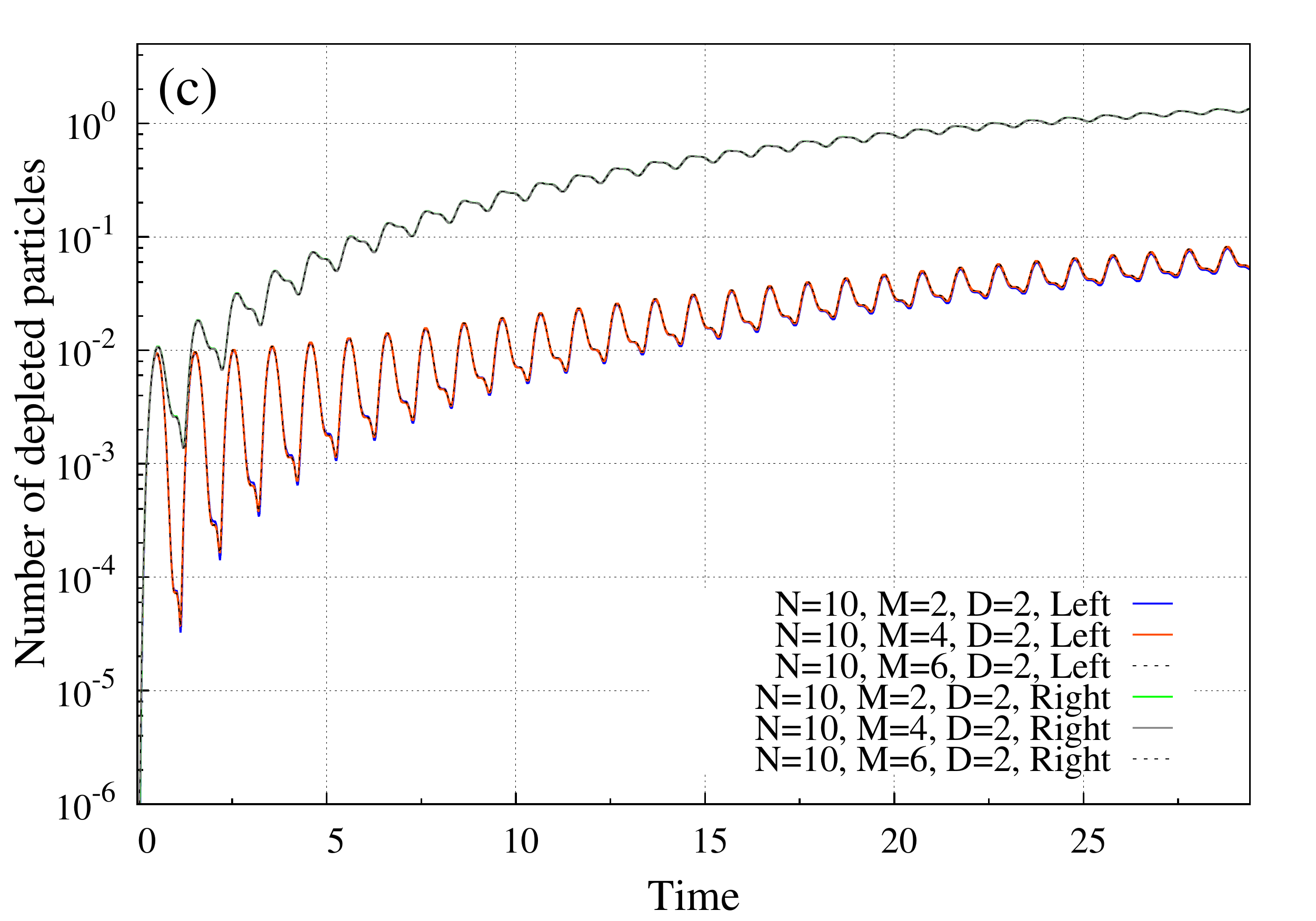} 
			\includegraphics[width=0.32\linewidth, angle=270]{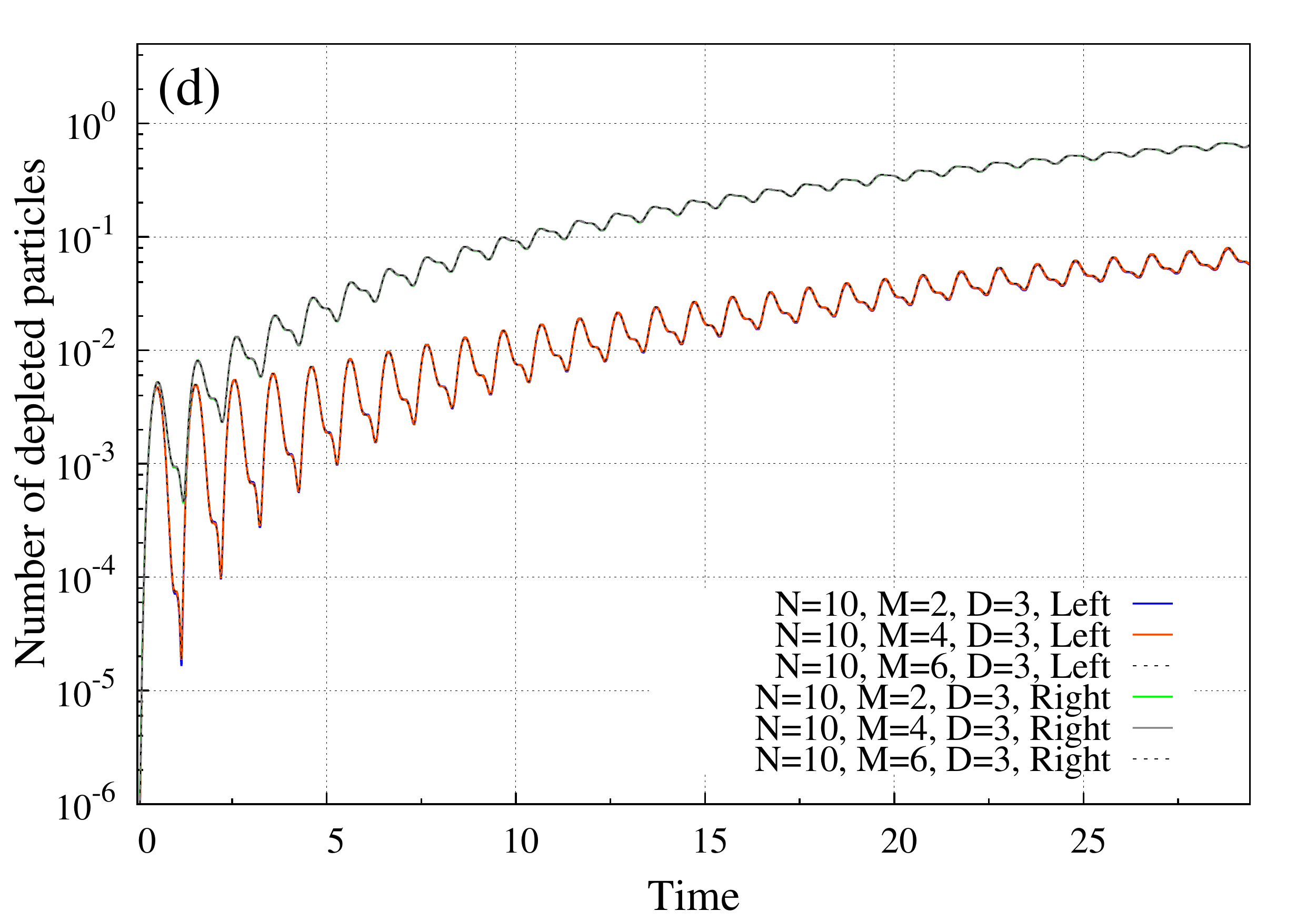} 
			\includegraphics[width=0.32\linewidth,angle=270]{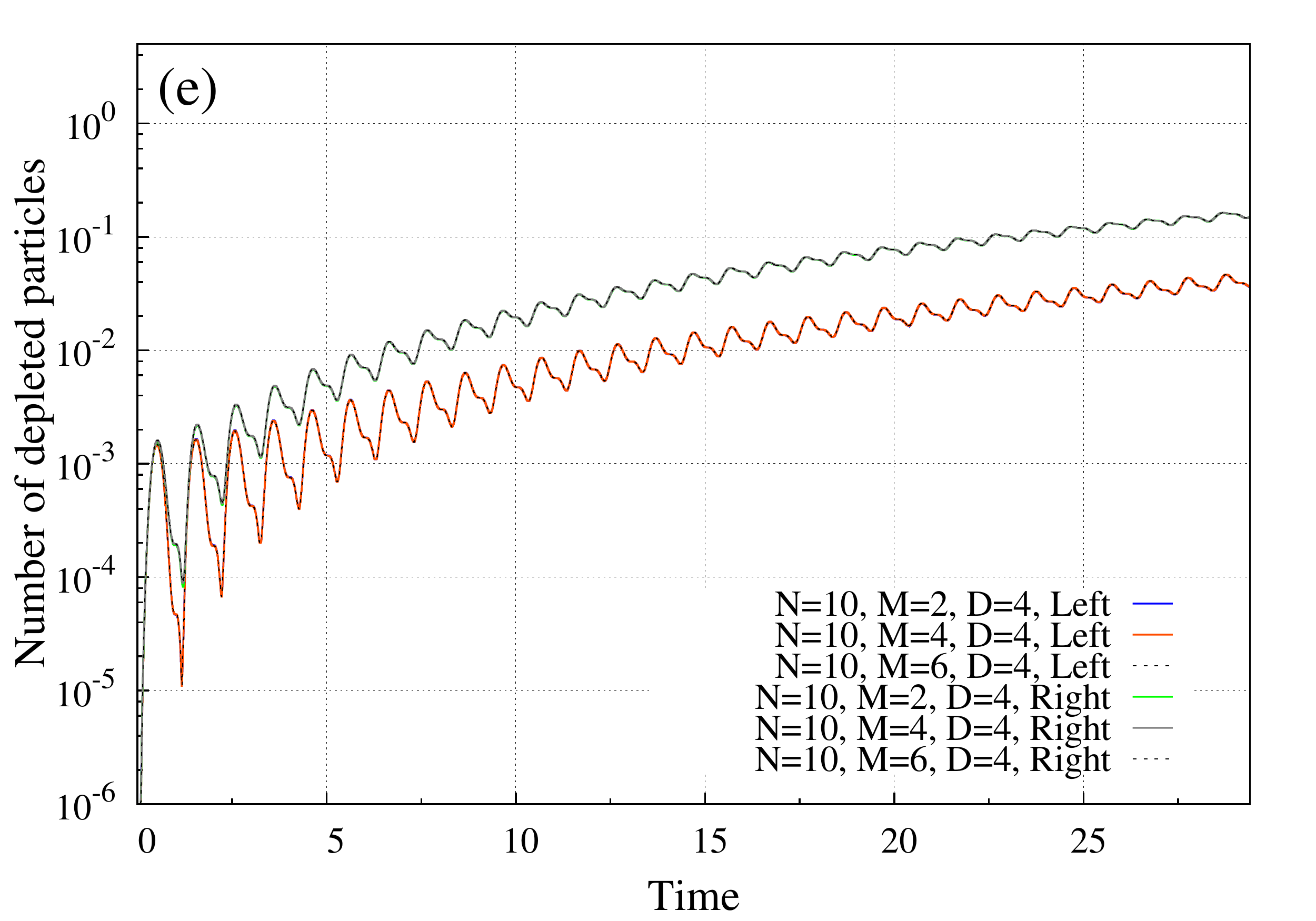} 
	\end{center}
	\caption{Convergence of the number of depleted atoms for both wells for (a) $D=l/8$, (b) $D=l/4$, (c) $D=l/2$, (d) $D=3l/4$ and (e) $D=l$. $l=4$ is the length scale of the system.
	Time is scaled by $t_{Rabi}$. See text for further details. The quantities shown here are dimensionless. Color codes are explained in each panel. }
	\label{fig5}
\end{figure}

Next, we plot the number of depleted atoms for each $D$ and both the left and right wells in each panel of Fig.~\ref{fig5}. Here also, we see that for all cases, 
the number of depleted atoms grow 
in an oscillatory manner. As discussed in the text, the system is more depleted for the right well than the left well for all $D$. 
Also, the depletion of the system is found to depend on the range $D$ of the interaction more prominently for the right well.
Here also, we find that the number of depleted particles computed with $M=2,4,$ and $6$ 
practically overlap with each other while the results for $M=4$ and $6$ are again lying on top of each other. This shows that the results obtained with $M=2$ orbitals are already
accurate and aptly describe the impact of the asymmetry of the trap and the range of the interaction at the many-body level.

\end{document}